\documentclass[aps,pre,reprint,showpacs,superscriptaddress]{revtex4-1}
\usepackage{graphicx}
\usepackage{dcolumn}
\usepackage{bm}
\usepackage{color}
\begin{document}


\title{Monte-Carlo approach to calculate the proton stopping in warm dense matter within particle-in-cell simulations}



\author{D. Wu}
\affiliation{State Key Laboratory of High Field Laser Physics, 
Shanghai Institute of Optics and Fine Mechanics, 201800 Shanghai, China}
\affiliation{Helmholtz Institut Jena, D-07743 Jena, Germany}
\author{X. T. He}
\affiliation{Key Laboratory of HEDP of the Ministry of Education, Center for Applied Physics and Technology, 
Peking University, 100871 Beijing, China}
\author{W. Yu}
\affiliation{State Key Laboratory of High Field Laser Physics, 
Shanghai Institute of Optics and Fine Mechanics, 201800 Shanghai, China}
\author{S. Fritzsche}
\affiliation{Helmholtz Institut Jena, D-07743 Jena, Germany}
\affiliation{Theoretisch-Physikalisches Institut, Friedrich-Schiller-University Jena, D-07743 Jena, Germany}

\date{\today}

\begin{abstract}   
A Monte-Carlo approach to proton stopping in warm dense matter is implemented into an existing particle-in-cell code. The model is based on multiple binary-collisions among electron-electron, electron-ion and ion-ion, taking into account contributions from both free and bound electrons, and allows to calculate particle stopping in much more natural manner. At low temperature limit, when ``all'' electron are bounded at the nucleus, the stopping power converges to the predictions of Bethe-Bloch theory, which shows good consistency with data provided by the NIST. With the rising of temperatures, more and more bound electron are ionized, thus giving rise to an increased stopping power to cold matter, which is consistent with the report of a recently experimental measurement [Phys. Rev. Lett. 114, 215002 (2015)]. When temperature is further increased, with ionizations reaching the maximum, lowered stopping power is observed, which is due to the suppression of collision frequency between projected proton beam and hot plasmas in the target.    

\end{abstract}


\maketitle

\section{Introduction}

In many cases of high energy density physics researches, the temperature of the target may reach $1$-$100\ \text{eV}$ while still maintaining the near-solid density, the original cold target becomes warm dense matter (WDM) \cite{wdm1,wdm2,wdm3}. Until now, however, the properties of WDM are not well understood. This is because both the methods for condensed-matter and for high-temperature plasmas are needed in order to well organize the intermediate regime of WDM.

The heating of the material used to be through lasers. In direct laser heating \cite{icf1,icf2,hedp.1,hedp.2}, the majority of the laser energy is absorbed by electrons at the front surface of the target, with the heating depth around $100\text{s}\ \text{nm}$. 
With the rapid progression of short-pulse and high-intensity laser technology, it has been demonstrated that highly energetic proton beams can be efficiently accelerated and focused from laser-irradiated solid targets \cite{lpi1, lpi2, lpi3, lpi4}.
Energetic, i.e., $\text{MeV}$, ion-beam-solid interactions provide a much more efficient heating mechanism, 
i.e., isochoric heating, which is usually of deep ($100\text{s}\ \mu\text{m}$) and localized energy depositions \cite{isochoric}. 
Thus, understanding the energetic ion beam dynamics in the matter is of importance for the wide range of potential applications, such as medicine physics including tumour therapy \cite{appli1}, creation of the WDM \cite{wdm1, wdm2, wdm3}, and the ion fast ignition concept of inertial confinement fusion \cite{lpi3}. 

In cold materials, the stopping of individual proton is already well understood. 
Bethe-Bloch theoretical approach \cite{b-b_1,b-b_2,b-b_3,b-b_4} is the basic method for evaluating the energy loss of protons with energies above $1\ \text{MeV}$. 
In high temperature plasmas, the long-range Coulomb-scattering is the well acknowledged framework 
for describing particle transportations \cite{jackson}.
Nevertheless, the proton stopping and energy deposition in WMD is still not clear. Until the present, however, there are few works focused on proton stopping in warm dense matter or partially ionized materials, in which both bound and free electrons contributes significantly. 
Recently, a high-accuracy measurement of charged-particle energy loss through WDM was reported \cite{warm_exp_1, warm_exp_2}, which shows an increased loss relative to cold matter. Almost simultaneously, hybrid particle-in-cell (PIC) simulations of intense proton beam's transport and energy deposition in solid-density matter were performed \cite{hybrid_pic_1,hybrid_pic_2}. However, in their research, the reduction of ``warm-target'' stopping power from the cold condition was reported.

In the present work, a Monte-Carlo approach within the framework of particle-in-cell (PIC) simulation is configured to proton stopping in warm dense matter. The model is based on multiple binary-collisions among electron-electron, electron-ion and ion-ion, which allows to calculate proton stopping in much more natural manner. 
Contributions from both bound and free electrons are considered and organized in a self-consistent way. 
Compared with existing models, ``all'' physical quantities, like angular scattering, momentum transferring and temperature variation, can be figured out under this approach. 
At low temperature limit, the stopping power calculated by the PIC code converges to the predictions of Bethe-Bloch theory, well consistent with the NIST. With the rising of temperatures, more and more bound electron are ionized, giving rise to an increased stopping power to cold matter. This increased energy loss relative to cold conditions supports the recent measurement in experiment \cite{warm_exp_1, warm_exp_2}. When temperature is further increased, with ionizations reaching the maximum, lowered stopping power is observed, which can be explained by the suppression of collision frequency between projected proton beam and hot plasma background.

The paper is organized as follows. 
In Sec. II, Bethe-Bloch and free electrons stopping power models are separately analysed, and self-consistently organized together and implemented into the PIC code. Then, contributions from bound and free electrons are investigated. 
In Sec. III, the results of PIC simulations are presented. 
The stopping powers at low temperature limit are calculated and compared with the NIST. Variations of stopping power and proton beam range with the increase of bulk temperature are also studied in this section. Summary and discussion are given in Sec. IV.   

\section{Physical model}

When an energetic (MeV) proton beam irradiates at a solid target, the proton will experience a friction force by colliding with electrons and nucleus of the target. As the stopping of protons mainly comes from the friction of electrons other than nucleus, which is usually $m_e/m_i$ times smaller, the contribution of nucleus is ignored in the present work. Furthermore, the contributions from electrons contain two parts, one is from bound electrons and the other is from free electrons. The fractions of bound and free electrons are determined by the ionization degree. Thus the total stopping power, following the commonly used approach \cite{hybrid_pic_1,hybrid_pic_2}, can be treated as the summation of the two contributions, 
i.e., ${(dE/dz)}_{\text{t}}={(dE/dz)}_{\text{b}}+{(dE/dz)}_{\text{f}}$, 
where the former one is the bound electrons' contribution and the latter one is the free electrons' contribution.  

For cold materials, when ``all'' electrons are bound at the nucleus, the energy loss of protons is primarily due to the processes of ionization and excitation, via the Coulomb force, of the electron cloud surrounding the nucleus. The Bethe-Bloch model accounts for both ionization and excitation of the atomic electrons and has the form \cite{theory_1,theory_2},
\begin{equation}
\label{B-B}
(\frac{dE}{dz})_{\text{b}}=\frac{4\pi e^2Z_{\text{A}}n_i}{m_e\beta^2}
[\ln{\frac{2\gamma m_e\beta^2}{\bar{I}}}-\beta^2-\frac{C_{\text{K}}}{Z_{\text{A}}}-\frac{\delta}{2}],
\end{equation}
where $Z_{\text{A}}$ is the atomic number of stopping medium, $n_i$ is the nucleus density of stopping medium, $m_e$ is the electron mass, $\gamma$ is the relativistic factor of the projected proton, $\beta$ is the velocity of projected proton, and $\bar{I}$ is the average ionization potential. The average ionization potential used in Eq.\ (\ref{B-B}) is formally defined by $Z\ln{\bar{I}}=\sum_n f_n\ln{E_n}$, where $E_n$ and $f_n$ are possible electronic energy transition and corresponding dipole oscillator strengths for the stopping medium. In practice, the oscillator strengths and transition energies are not well known. Rather experiments are done to empirically determine a value \cite{theory_1, theory_2}, i.e., $\bar{I}=11.5 \times Z_{\text{A}}\ \text{eV}$.
Note that in Eq.\ (\ref{B-B}), two additional corrective terms are included, the shell correction term, $C_{\text{K}}/Z_{\text{A}}$, and the density effect correction term, $\delta/2$. These two terms are based on Fano's original work \cite{b-b_5}. For shell correction $C_{\text{K}}$, it is a function of the quantity $\xi\equiv(c^2/\beta^2)Z_{\text{A}}^2\alpha^2$, 
which represents the squared ratio of the K-shell velocity to the projected proton velocity, 
where $\alpha$ is the fine structure constant. 
A simple approximation form for $C_{\text{K}}$ is $C_{\text{K}}=2.3\xi/(1.+1.3\xi^2)$. For density effect correction $\delta/2$, 
there is usually no simple relationship between its magnitude and atomic number of the stopping medium. 
Fortunately, $\delta/2$ have already been tabulated for all elemental targets \cite{delta_table}.

For warm dense matter, the Bethe-Bloch equation needs to be scaled with the material ionization. Particularly, the average ionization potential is the most important parameter. It is indicated that $\bar{I}$ should vary from the neutral atom value up to a value given by $I_0Z_{\text{A}}^2$ in the limit of a single K-shell electron remaining bound, where $I_0$ is the average ionization potential for hydrogen, i.e., $13.6\ \text{eV}$. By further assuming that the energy transitions scale as $Z^2$ and the dipole oscillator strengths remain constant, a scaling of the average ionization potential with ionization degree $Z$ can be obtained, of the form \cite{theory_1}, $\bar{I}(Z)=\bar{I}(Z_{\text{A}}-Z)Z_{\text{A}}^2/(Z_{\text{A}}-Z)^2$. This approximate expression has the desirable features that it reproduces the single K-shell electron limit and also includes the shell structure variations in the neutral atom average ionization potentials.

To describe the free electron stopping power, multiple binary collision models with a Debye radius coupled with collective plasma wave excitation outside the Debye radius is normally considered. 
Following Jackson \cite{jackson}, we can write the free electron stopping power as, 
\begin{equation}
\label{free}
(\frac{dE}{dz})_{\text{f}}=\frac{4\pi e^2Zn_i}{m_e\beta^2}G(\eta)\ln(\Lambda_{\text{f}}),
\end{equation}  
where $Z$ is the ionization charge state, $G(\eta)=\text{erf}(\sqrt{\eta})-2\sqrt{\eta/\pi}\exp(-\eta)$ and $\eta=m_ec^2\beta^2/2T_e$. 
For the Coulomb logarithm, $\ln{(\Lambda_{\text{f}})}$, it is usually defined as $L\equiv\ln(1/\theta_{\min})$, 
where $\theta_{\min}$ is the smallest angle for which the process can still be regarded as small angle Coulomb scattering. For classical scattering, i.e., $1\ll2\pi e^2/h \beta$, we have $L=\ln{(\lambda_{\text{D}}m_e\beta^2/e^2)}$, here $h$ is the Plank constant and $\lambda_{\text{D}}$ is the Debye length. This condition is not satisfied in the relativistic case, so that the scattering must be treated quantum-mechanically using the Born approximation. In this case, then, $L=\ln{(2\pi\lambda_{\text{D}}\gamma \beta/h)}$ is expressed as the ratio of the Debye length and the de Broglie wave length.     

It is well recognized that electron-electron, electron-ion and ion-ion scatterings in plasmas can be well processed by Monte-Carlo binary collision model. Following the pioneering works of Takizuka \cite{pic_collision_1}, Nanbu \cite{pic_collision_2} and Sentoku \cite{pic_collision_3,pic_collision_4}, fully relativistic energy-conserving Monte-Carlo binary collision models have been implemented into the particle-in-cell simulation code, which is the newly extended version based on LAPINE \cite{code1}. Within the PIC calculations, i) pairs of particles, electron-electron, electron-ion or ion-ion, suffering binary collisions are determined at random in a cell, ii) the changes in the velocity of two particles due to a binary collision in the time interval $\delta t$ are computed, iii) and then the velocity of each particle is replaced by the newly calculated one. Note that the Monte-Carlo binary collision model, in principle, can not account for bound electrons, since they are already present in the ions. However in warm dense matter or partially ionized material regime, both bound and free electrons contribute significantly to the stopping of projected protons. Fortunately, when comparing Eq.\ (\ref{B-B}) and Eq.\ (\ref{free}), we find that the stopping power of bound and free electrons share the same behaviours, except the different Coulomb logarithms. To include bound electrons' contribution in the binary collision model,
we revise the electron-ion scattering term (or ion-electron collision frequency), in the above ii) step, as,
\begin{equation}
\label{scattering}
\nu_{\text{i-e}}=\frac{8\sqrt{2\pi} e^4Zn_i}{3\gamma^2\beta^3}[\ln{(\Lambda_{\text{f}})}+\frac{Z_{\text{A}}-Z}{Z}\ln{(\Lambda_{\text{b}})}],
\end{equation}   
where $\ln{(\Lambda_{\text{b}})}\equiv\ln{({2\gamma m_e\beta^2}/{\bar{I}})}-\beta^2-{C_{\text{K}}}/{Z_{\text{A}}}-{\delta}/{2}$,
and ${(Z_{\text{A}}-Z)}/{Z}$ defines the ratio of bound electrons' contribution. 
When $Z\rightarrow Z_{\text{A}}$, i.e., fully ionized plasmas, 
electron-ion scattering term in Eq.\ (\ref{scattering}) 
converges to $\nu\sim[{8\sqrt{2\pi} e^4Zn_i}/{3\gamma^2\beta^3}]\ln{(\Lambda_{\text{f}})}$. This is exactly the cases for pure plasmas. When $Z\rightarrow 0$, i.e., neutral atoms, electron-ion scattering term in Eq.\ (\ref{scattering}) is $\nu\sim[{8\sqrt{2\pi} e^4Z_{\text{A}}n_i}/{3\gamma^2\beta^3}]\ln{(\Lambda_{\text{b}})}$. 

Our model, in principle, is only suitable for proton transportation with moderate and high energy, i.e. MeV-GeV levels. This is because, even with the inclusion of shell correction, the Bethe-Bloch theory is not appropriate for very low energy ions. In this regime, Linhard-Scharff-Schiott model \cite{j-s-s}, instead, needs to be used. However this correction is not included in our present work. To avoid the divergence of collision frequency, i.e., when $\beta\rightarrow 0$ in Eq.\ (\ref{scattering}), we have set a threshold of collision frequency, i.e., $\nu=(4m_eZe^4/3\pi h^3)L$ \cite{degenerate}, which is the collision frequency limit in the degenerate regime. 

\begin{table}
\caption{\label{table1} Coulomb logarithm as a function of energy of projected protons for solid aluminium 
at temperature $T_e=150\ \text{eV}$. Both $\Lambda_{\text{f}}$ and $\Lambda_{\text{b}}$ are calculated in the PIC code by averaging over $1000$ projected protons.}
\begin{ruledtabular}
\begin{tabular}{ l  l  l  l  l  l  l  l }
MeV & 0.10 & 0.50 & 1.0 & 5.0 & 10.0 & 50.0 & 100.0 \\
$\Lambda_{\text{f}}$ & 10.52 & 11.11 & 11.56 & 12.2 & 12.88 & 13.58 & 13.99 \\ 
$\Lambda_{\text{b}}$ & 0.21 & 1.33 & 2.17 & 2.97 & 4.24 & 5.58 & 6.32 \\
\end{tabular}
\end{ruledtabular}
\end{table}

According to Eq.\ (\ref{scattering}), ionization degree is one of the the dominant quantities that determine contributions from bound and free electrons. In our previous work, we have established a physical model \cite{eos0}, in which the ionization dynamics of warm dense matter can be self-consistently calculated. Thus, the ionization distributions in each computational cell can be evaluated in the PIC simulation, and then the average ionization degree $Z$ of each species can be determined. Following Eq.\ (\ref{scattering}), 
the total Coulomb logarithm $\ln{(\Lambda_{\text{t}})}\equiv\ln{(\Lambda_{\text{f}})}+[{Z_{\text{A}}-Z)}/{Z}]\ln{(\Lambda_{\text{b}})}$ is updated at each computational cell per time step. As an example, for aluminium target of density $2.7\ \text{g}/\text{cm}^2$ and temperature $150\ \text{eV}$, the corresponding ionization degree is calculated to be $Z=5$. Table.\ \ref{table1} lists values of $\Lambda_{\text{f}}$ and $\Lambda_{\text{b}}$ as functions of projected proton energies, calculated in the PIC code by averaging over $1000$ projected protons. It is indicated that $\Lambda_{\text{b}}$ is smaller than $\Lambda_{\text{f}}$ for all energy ranges, and that both $\Lambda_{\text{f}}$ and $\Lambda_{\text{b}}$ increase with the increase of projected proton energy, although the variations of $\Lambda_{\text{f}}$ are relatively small.   

\section{Applications}
\begin{figure}
\includegraphics[width=8.50cm]{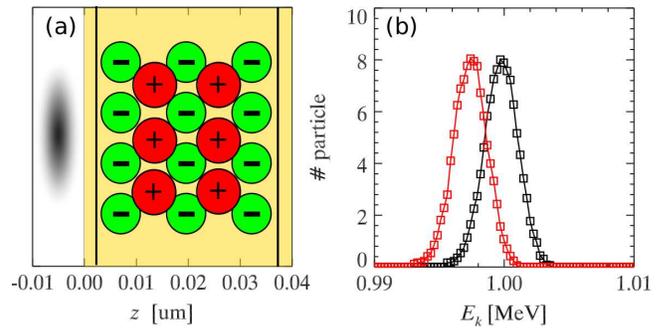}
\caption{\label{fig1} (color online) (a) Configuration of proton stopping calculation in our PIC simulations. A proton beam of energy $1\ \text{MeV}$ irradiates the aluminium bulk. Two diagnostic planes separated by a distance of $0.039\ \mu\text{m}$ are placed to record the energy spectra of protons. (b) black-square line represents the spectra recorded by the first diagnostic plane, while the red-square line represents that by the second one.}
\end{figure}

The first we need to do is to check the stopping power calculations by the PIC code. Our calculation configuration is shown in Fig.\ \ref{fig1}. As shown in Fig.\ \ref{fig1} (a), the size of the simulation box is choose to be $0.04\ \mu\text{m}$, which is divided into $400$ cells. Simulation time step is $\delta t =0.0003\ \text{fs}$. Two diagnostic planes separated by a distance $0.039\ \mu\text{m}$ are located at $z_1=0.0005\ \mu\text{m}$ and $z_2=0.0395\ \mu\text{m}$.  
A proton beam of energy $1.0\ \text{MeV}\pm1.0\ \text{eV}$ is projected from the left boundary. The energy spectra recorded by these two diagnostic planes are shown in Fig.\ \ref{fig1} (b), with black-squared-line for proton spectra recorded at $z_1=0.005\ \mu\text{m}$ and red-squared-line at $z_2=0.0395\ \mu\text{m}$. The stopping power can then be calculated as the peak energy shift divided by the distance of the two diagnostic planes. Note that in the PIC simulation, each binary collision is assumed to be small angle. As the scattering angle usually satisfy a Gaussian distribution, i.e., $<\tan^2(\theta/2)>=\nu \delta t$, $\delta t$ should be very small. To ensure the accuracy of our calculation, we decreased $\delta t$ by two and five times, and found no significant difference of the final results.    

\begin{figure}
\includegraphics[width=8.50cm]{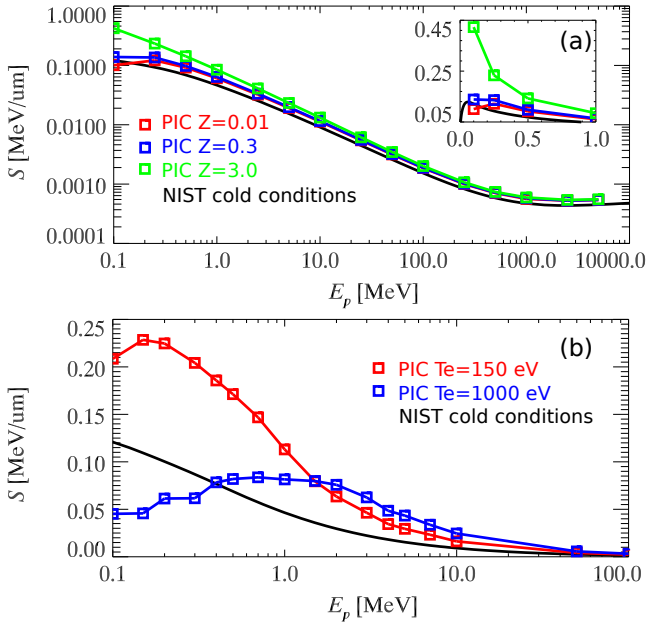}
\caption{\label{fig2} (color online) Stopping power as a function of projected proton energy. Results from our PIC simulations are compared with that from the NIST. (a) Results of PIC simulations: red-square line is the one provided the average ionization degree is $\bar{Z}=0.01$, blue-square line is the one provided the average ionization degree is $\bar{Z}=0.3$, and green-square line is the one with average ionization degree of $\bar{Z}=3.0$. Data from the NIST is plotted in black solid line. 
(b) Results of PIC simulations: Red-square line is the one provided $T_e=150\ \text{eV}$ with average ionization degree $\bar{Z}=5$, and blue-square line is the one provided $T_e=1000\ \text{eV}$ with average ionization degree $\bar{Z}=11$. 
Data from the NIST is plotted in black solid line.
Note the average ionization degree under different temperature is calculated based on either the model we proposed in reference \cite{eos0} or the external EOS tables \cite{eos1,eos2}.}
\end{figure}

Following the routine described above, we compared the stopping power calculated by the PIC code with that from the NIST. As the values of stopping power from NIST are measured/calculated at room temperatures, in our calculation, the temperature of the target is also set to be very low, which is $1.0\ \text{eV}$. Fig.\ \ref{fig2} (a) shows the stopping power of protons in solid aluminium as a function of projected proton energy, from $0.1\ \text{MeV}$ to $10000.\ \text{MeV}$. The black curve in Fig.\ \ref{fig2} (a) represents the stopping power values from the NIST \cite{nist}. The red-squared-line, blue-squared-line and green-squared-line represent the calculated stopping powers by PIC code, provided that the initial ionization degrees are $\bar{Z}=0.01$ (red-squared-line), $\bar{Z}=0.3$ (red-squared-line) and $\bar{Z}=3$ (green-squared-line). As shown in Fig.\ \ref{fig2} (a), the PIC calculated stopping powers converge to the NIST with the average ionization degree approaching to zero at low temperature limit. Through these comparisons, it is reasonable to claim that i) the physical model implemented in the PIC code is benchmarked and makes sense, 
ii) at room temperature the average ionization degree of aluminium should be close to zero. However, it is $\bar{Z}=3$ that is assumed in a recent hybrid PIC simulation \cite{hybrid_pic_1,hybrid_pic_2}.

\begin{figure}
\includegraphics[width=8.50cm]{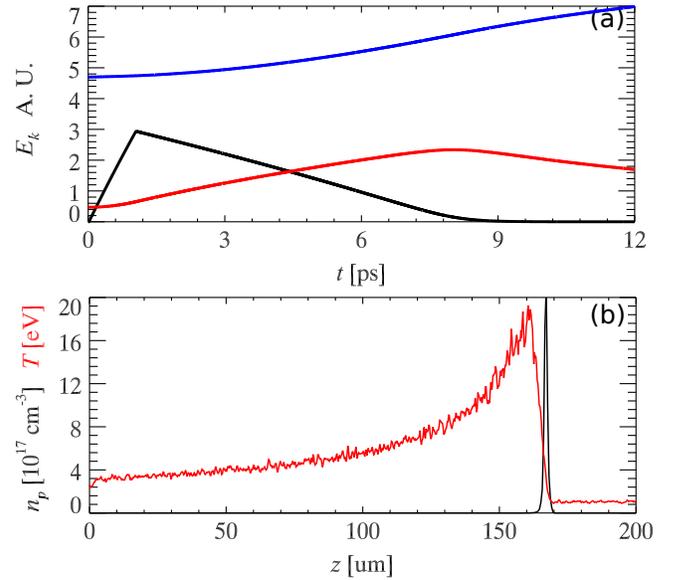}
\caption{\label{fig3} (color online) (a) Particle kinetic energies as function of time in a typical PIC simulation. 
A proton beam, with energy of $5\ \text{MeV}$, density of $10^{17}\ \text{cm}^{-3}$ and pulse duration $1\ \text{ps}$, irradiates the bulk aluminium, with initial temperature $T_e=1\ \text{eV}$ and average ionization degree $\bar{Z}=0.1$. 
Black line represents the energy of the injected protons, blue line is the kinetic energy of ions (excluding the projected protons) in the simulation box, and red line is the total electron energy in the simulation box. 
(b) Proton density (black) distribution and electron background temperature (red) at the final time of simulations.}
\end{figure}

\begin{figure}
\includegraphics[width=8.50cm]{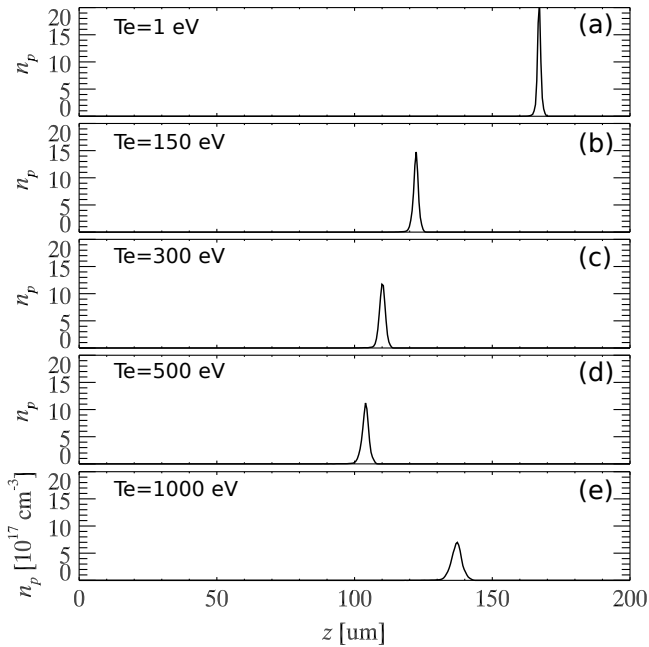}
\caption{\label{fig4} (color online) Variation of proton beam range with the increase of bulk temperatures. 
A proton beam, with energy of $5\ \text{MeV}$, density of $10^{17}\ \text{cm}^{-3}$ and pulse duration $1\ \text{ps}$, irradiates the bulk aluminium with different initial temperatures. 
(a) temperature $T_e=1\ \text{eV}$ with average ionization degree $\bar{Z}=0.1$, 
(b) temperature $T_e=150\ \text{eV}$ with $\bar{Z}=5$, 
(c) temperature $T_e=300\ \text{eV}$ with $\bar{Z}=8$, 
(d) temperature $T_e=500\ \text{eV}$ with $\bar{Z}=11$, 
and (e) temperature $T_e=1000\ \text{eV}$ with $\bar{Z}=11$.}
\end{figure}

When rising the temperature of the aluminium bulk, more and more bound electrons are ionized. At $T_e=150\ \text{eV}$, the average ionization degree is of $\bar{Z}=5$, while at $T_e=1000\ \text{eV}$, the average ionization degree is of $\bar{Z}=11$, which can be obtained either by our previously proposed calculation model \cite{eos0} or from other external EOS tables \cite{eos1, eos2}. The red-squared-line in Fig.\ \ref{fig2} (a) represents the stopping power of solid aluminium at $T_e=150\ \text{eV}$ by the PIC code. Compared with that at low temperature limit, an significant increase is observed. This is consistent with a recent experimental report \cite{warm_exp_1}. In this experiment, the stopping of energetic ($15\ \text{MeV}$) protons in an isochorically heated solid-density Be plasma with an electron temperature of $32\ \text{eV}$ was measured, which showed an increased stopping power to cold matter. This is because, as shown in Table.\ \ref{table1}, $\Lambda_{\text{f}}$ of the individual free electron is larger than $\Lambda_{\text{b}}$ of the individual bound electron at temperatures of $100\text{s}\ \text{eV}$. With the increase of the temperature, more free electrons contribute to stopping while the number of bound electrons significantly drops. However, when temperature reaches $T_e=1000\ \text{eV}$, as blue-squared-line in Fig.\ \ref{fig2} (a) shows, the decrease of stopping power is observed. This is because, the ionization has already reached the maximum, the rising of the temperature will no longer ionize more free electrons, but instead will depress the collision frequency of ion-electrons. From Eq.\ (\ref{scattering}), the ion-electron collision frequency depends on $\nu\sim1/(\gamma^2\beta^3)$. The rising of plasma temperature leads to the increase of $\beta$, which is the relative velocity between the projected proton and the electron in target. Furthermore, according to Eq.\ (\ref{free}), the free electron stopping power has a peak value when the projected proton velocity is near the electron thermal velocity in the target, this feature is also figured out by the PIC simulations, as shown in Fig.\ \ref{fig2} (b).     

Fig.\ \ref{fig3} shows the full process of energetic proton beam's stopping and transportation in solid aluminium calculated by PIC code. A proton beam with energy of $5\ \text{MeV}$, density of $10^{17}\ \text{cm}^{-3}$ and pulse duration $1\ \text{ps}$, irradiates the bulk aluminium, whose initial temperature and average ionization degree are assumed to be $T_e=1\ \text{eV}$ and $\bar{Z}=0.1$. The energy history is shown in Fig.\ \ref{fig3} (a). Black line represents the energy of the injected protons, blue line represents the kinetic energy of ions (excluding the projected protons) in the simulation box, and red line presents the total electron energy in the simulation box. At the very beginning, within the first $1\ \text{ps}$, the entire proton beam is projected into the bulk target. Then the protons start to lose their energies rapidly by transferring energy to electrons in the target. At time of $9\ \text{ps}$, almost all protons are stopped by donating all their kinetic energies. We also observe the gradual energy transfer from heated electron to ions in the target, which is determined by electron-ion collisions. Fig.\ \ref{fig3} (b) shows the ``equilibrium'' proton distribution and electron background temperatures when all protons are stopped. The bulk target is significantly heated along the path of projected proton beam, 
with a typical temperature $5\ \text{eV}$ and maximal temperature of $20\ \text{eV}$ at the end of the path. The range of the $5\ \text{MeV}$ proton beam in solid aluminium is $170\ \mu\text{m}$, 
consistent with the recent hybrid PIC simulation \cite{hybrid_pic_1}.

Furthermore, we also changes the target temperature to see how the proton beam range changes with the target temperatures. Fig.\ \ref{fig4} shows the variation of the proton beam range with the increase of bulk temperature. It is shown that the range firstly decreased and then increased, though the beam is broadened with the increase of target temperature. The variation of the proton beam range with bulk temperature can be easily explained if we follow the stopping power calculation shown in Fig.\ \ref{fig2}. 
Note that in the recent hybrid PIC simulation \cite{hybrid_pic_1, hybrid_pic_2}, only half of the physics is figured out, i.e., the proton beam range is increasing with the increase of background temperature. This difference is caused by the different stopping power and ionization calculation models, so it would be helpful to check which model is more reliable in future experiments.

\section{Discussion and conclusions}

To summarize, a physical model for proton stopping in warm dense matter or partially ionized materials is proposed and implemented into an existing particle-in-cell code. This model is based on Monte-Carlo binary collisions, which has the advantage to calculate proton stopping in much more natural manner. The model concerns binary-collisions among electron-electron, electron-ion and ion-ion, taking into account contributions from both free and bound electrons. 

At low temperature limit, when ``all'' electron are bounded at the nucleus, the stopping power converges to the predictions of Bethe-Bloch theory, and shows good consistency with data provided by the NIST. Through the comparisons with NIST, the correctness of physical model is benchmarked. Besides that, we also claim from these comparisons that at room temperature the average ionization degree of aluminium should be close to zero.

With the increase of temperature, more and more bound electron are ionized, giving rise to an increased stopping power to cold matter. This observation is consistent with the report of a recently experimental measurement. When temperature is further increased, with ionizations reaching the maximum, lowered stopping power is observed, which is due to the suppression of collision frequency between projected proton beam and hot plasmas target. 

The full process of proton beam's stopping and transportation in solid target is studied by PIC simulations. The variation of the proton beam range with target temperature is figured out. Results show that with the rising of the target temperature, the range firstly decreased and then increased. This finding is different from the report of a recent hybrid PIC simulation, in which only half of the physics is figured out, i.e., the proton beam range is increasing with the increase of background temperature. The difference is due to the different stopping power and ionization calculation models, to check which model is more reliable needs further experiment measurements in the future.

The multi-dimensional, nucleus scattering and self-generated electromagnetic field effects shall also be studied by this PIC code. The related results shall be presented in a following separated paper.

\begin{acknowledgments}
D. Wu wishes to acknowledge the financial support from German Academic Exchange Service (DAAD) and China Scholarship Council (CSC),
also thanks J. W. Wang and S. Z. Wu at Helmholtz Institut-Jena and S. X. Luan at Shanghai Institute of Optics and Fine Mechanics for fruitful discussions. 
\end{acknowledgments}

{}

\end{document}